\documentclass[preprint]{aastex}
\usepackage{graphicx}
\usepackage{array}
\usepackage{longtable}
\usepackage{natbib}
\usepackage{amssymb}
\begin{document}

\title{\textit{Object-X}: The Brightest Mid-IR Point Source in M33}
\author{Rubab~Khan\altaffilmark{1},
K.~Z.~Stanek\altaffilmark{1,2},
C.~S.~Kochanek\altaffilmark{1,2},
A.~Z.~Bonanos\altaffilmark{3}
}

\altaffiltext{1}{Dept.\ of Astronomy, The Ohio State University, 140
W.\ 18th Ave., Columbus, OH 43210; khan, kstanek, ckochanek@astronomy.ohio-state.edu}

\altaffiltext{2}{Center for Cosmology and AstroParticle Physics, 
The Ohio State University, 191 W.\ Woodruff Ave., Columbus, OH 43210}

\altaffiltext{3}{Institute of Astronomy \& Astrophysics, National
  Observatory of Athens, I. Metaxa \& Vas. Pavlou St., P. Penteli, 15236
  Athens, Greece; bonanos@astro.noa.gr}

\shorttitle{The Brightest Mid-IR Source in M33}

\shortauthors{Khan et al.~2010}
 
\begin{abstract}
\label{sec:abstract}

We discuss the nature of the brightest mid-IR point source (which we dub 
Object~X) in the nearby galaxy M33. Although multi-wavelength data on this 
object have existed in the literature for some time, it has not previously 
been recognized as the most luminous mid-IR object in M33 because it is 
entirely unremarkable in both optical and near-IR light. In the Local Group
Galaxies Survey, Object~X is a faint red source visible in $VRI$ and H$\alpha$ 
but not $U$ or $B$. It was easily seen at $JHK_s$ in the 2MASS survey. It is 
the brightest point source in all four \textit{Spitzer} IRAC bands and is also visible 
in the MIPS 24~$\micron$ band. Its bolometric luminosity is $\sim5\times10^5 L_\odot$.
The source is optically variable on short time 
scales (tens of days) and is also slightly variable in the mid-IR, indicating 
that it is a star. Archival photographic plates (from 1949 and 1991) 
show no optical source, so the star has been obscured for at least half a 
century. Its properties are similar to those of the Galactic OH/IR star IRC+10420 
which has a complex dusty circumstellar structure resulting from episodic low 
velocity mass ejections. We propose that Object~X is a $M\gtrsim 30 M_{\odot}$ 
evolved star obscured in its own dust ejected during episodic mass loss events 
over at least $\sim$half a century. It may emerge from its current ultra-short 
evolutionary phase as a hotter post-RSG star analogous to M33 Var~A. The 
existence and rarity of such objects can be an important probe of a very brief 
yet eventful stellar evolutionary phase.
 
\end{abstract} 
\keywords{stars: evolution, mass-loss, winds, outflows
--- galaxies: individual (M33)}
\maketitle

\section{Introduction}
\label{sec:introduction}

Explaining mass loss from massive stars, especially episodic mass loss in evolved massive 
stars, is one of the outstanding problems in stellar evolution 
theory. Systematic studies of rare, luminous, dusty massive stars in nearby 
galaxies were carried out in parallel by \cite{ref:Bonanos_2009,ref:Bonanos_2010} for the 
Large and Small Magellanic Clouds (LMC and SMC), and \cite{ref:Thompson_2009} and \cite{ref:Khan_2010a} 
for M33, NGC~300, M81, and NGC~6946. Luminous blue 
variables (LBVs), supergiant B[e] (sgB[e]), some Wolf-Rayet stars (WRs) and red 
supergiants (RSGs) contribute to this class of objects, and their rarity 
implies they are a very short but perhaps critical stage in the evolution of 
massive stars. 
Given that the LBV mass loss mechanism is poorly understood~\citep{ref:Humphreys_1994,ref:Smith_2006} and 
that there is mounting evidence for some core collapse supernovae with dust 
enshrouded progenitors~\citep{ref:Prieto_2008a,ref:Thompson_2009} or very 
recent mass loss episodes~\citep[][and references therein]{ref:GalYam_2007,ref:Smith_2008}, a census of these stars in nearby galaxies is 
vital for understanding the mass loss mechanisms and final stages of evolution 
of massive stars.

\cite{ref:Bonanos_2009,ref:Bonanos_2010} cross-matched massive stars with known 
spectral types with the SAGE and SAGE-SMC photometric databases, which 
resulted in a multi-band photometric catalog from 0.3 to 24$\mu$m 
providing spectral energy distributions (SEDs) for most 
classes of hot and cool stars. These works 
showed that in the LMC and the SMC, the LBVs, RSGs, and sgB[e] stars are some of the 
most-luminous mid-IR sources, due to the combination of high actual luminosities with 
surrounding dust from recent episodic mass ejections. 
However, these studies required the presence of a bright optical, spectroscopically classified source. 
The opposite approach, focusing on bright mid-IR sources, was undertaken by 
\cite{ref:Thompson_2009} and \cite{ref:Khan_2010a}.

The \cite{ref:Thompson_2009} re-analysis of the \cite{ref:McQuinn_2007} mid-IR images of M33 focused on 
detecting extremely red (optically thick even at 3.6$\mu$m) analogs of the 
progenitors of SN~2008S and the NGC~300 optical transient, finding that such analogs 
are extremely rare. Less red, but more luminous, 
stars are still rare, and only in some cases do they correspond 
to LBVs and other spectroscopically classified evolved stars. In 
\cite{ref:Khan_2010a}, we carried out a systematic mid-IR photometric search for 
self-obscured stars in four galaxies: M33, NGC~300, M81, and NGC~6946. In particular, we 
confirmed the conclusion of~\cite{ref:Thompson_2009} that stars analogous to the 
progenitors of SN~2008S and NGC~300 transients are truly rare---there may
be as few as $\sim1$ per galaxy at any given moment. This result empirically 
supports the idea that the dust-enshrouded phase is a very short-lived 
phenomenon in the lives of some massive stars.

\begin{figure*}[t]
\begin{center}
\includegraphics[width=120mm]{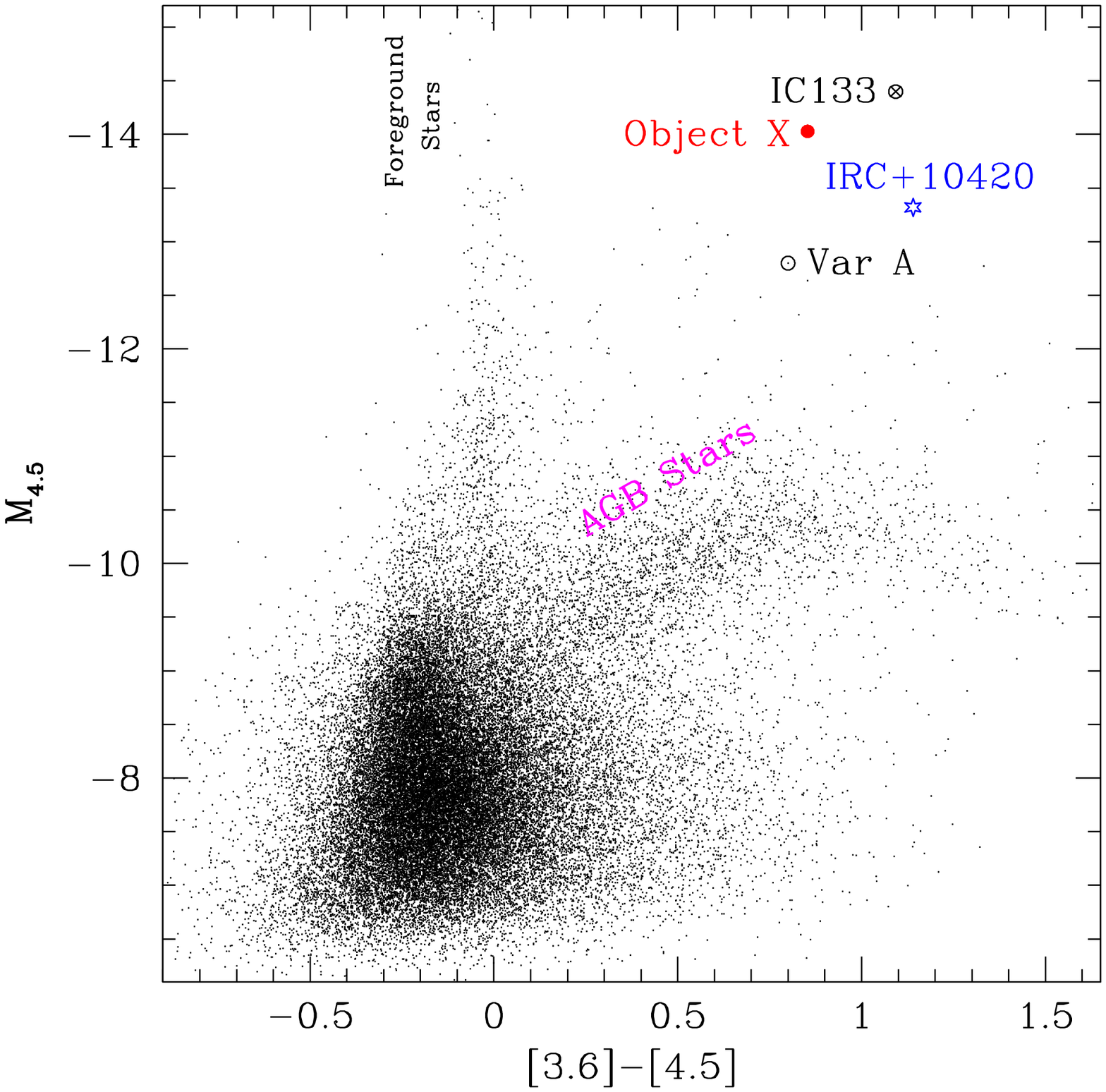}
\end{center}
\caption{The 4.5~$\micron$ absolute magnitude vs. the $[3.6]-[4.5]$ color mid-IR CMD of M33 adopted from~\cite{ref:Thompson_2009}. Object~X (red circle), the star cluster IC133 (cross and circle) and Var~A (open circle around dots) are marked. For comparison, we also show the position of IRC+10420 (blue starred symbol) for a 5~kpc distance~\citep{ref:Jones_1993}. The extremely bright stars near color $\sim0$ are the foreground stars and the stars on the red branch at $M_{4.5}\simeq-10.5$ are M33 AGB stars. There are no other stars in M33 that are even remotely similar to Object~X in these bands.}
\label{fig:m33_CMD1}
\end{figure*}

\begin{figure*}[ht]
\begin{center}
\includegraphics[width=120mm]{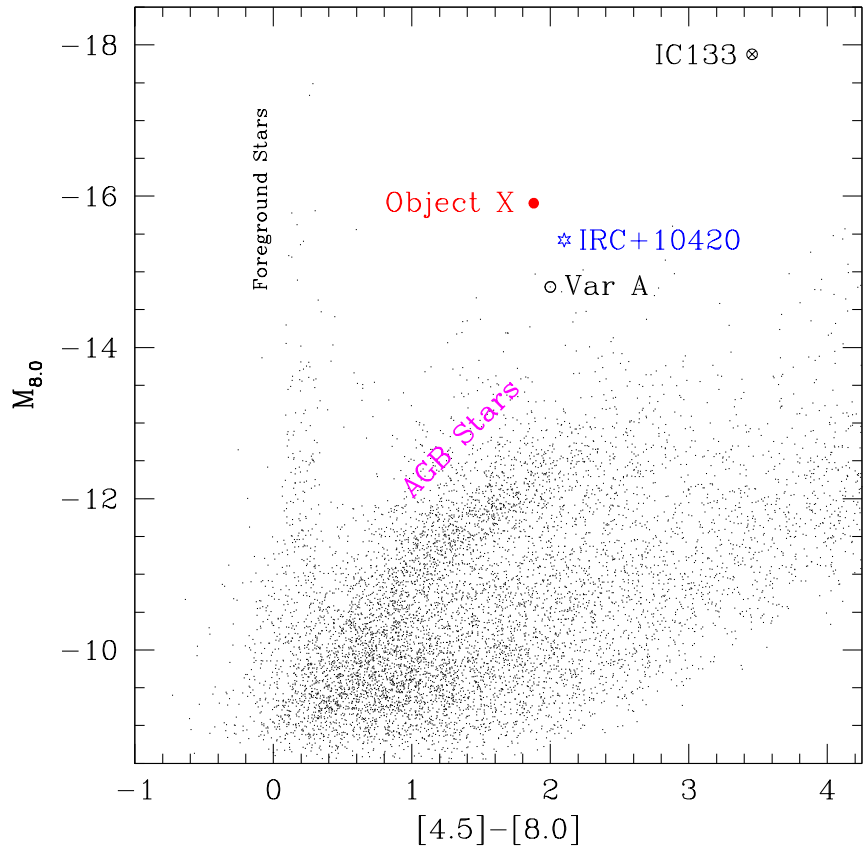}
\end{center}
\caption{Same as Figure \ref{fig:m33_CMD1} but for 8.0~$\micron$ absolute magnitude vs. the $[4.5]-[8.0]$ color. Object~X stands out in this combination of mid-IR bands as well.}
\label{fig:m33_CMD2}
\end{figure*}

At this point, we set out to examine the most luminous mid-IR sources 
in M33 that had not been considered by either \cite{ref:Thompson_2009} or 
\cite{ref:Khan_2010a}. In this paper, we discuss the nature of the brightest 
mid-IR star in M33. We started by examining the 
brightest objects off the primary stellar locus in the mid-IR CMD of 
M33~\citep[Figure~\ref{fig:m33_CMD1}, adopted from][also see Figure~\ref{fig:m33_CMD2}]{ref:Thompson_2009}. 
The brightest source is a known compact, young star cluster IC133. It is 
spatially unresolved by \textit{Spitzer}, but immediately recognizable as a star 
cluster. It is both too luminous to be a single star ($L_\star>10^7L_\odot$) 
and its mid-IR SED peaks at a temperature ($\sim30$K) that is too cold for dust associated 
with material ejected by a single star (because the required mass $\propto L_\star T_d^{-4}$).
The second brightest source, the brightest mid-IR star in all 
of M33, is the subject of this paper. We call this star, located at
R.A.=1$^h$33$^m$24\fs1 and Dec.=+30\arcdeg25\arcmin34\farcs8 (J2000.0), ``Object~X''. 

\begin{figure*}[t]
\begin{center}
\plotone{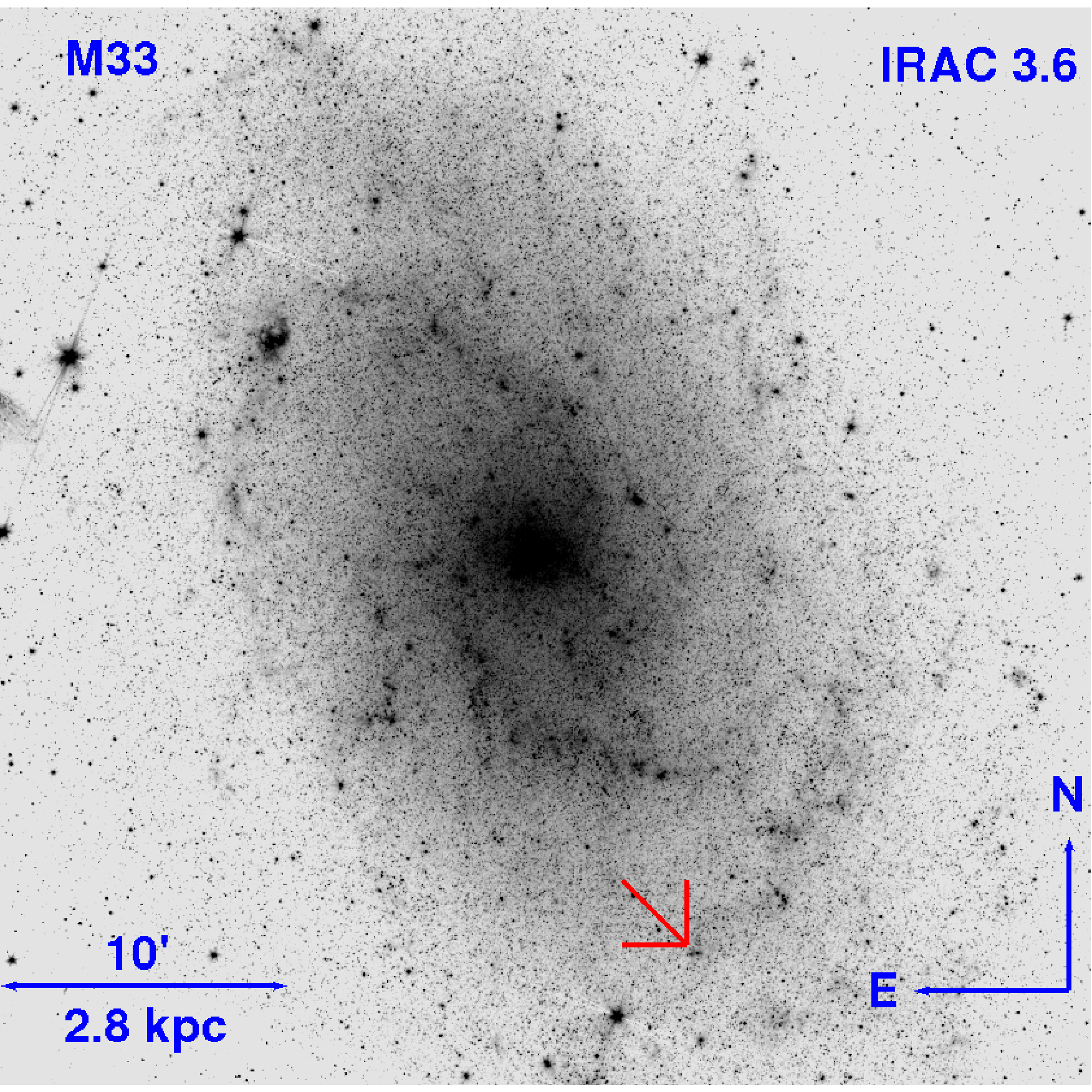}
\end{center}
\caption{The IRAC 3.6~$\micron$ image of M33 showing the location of Object~X 
(R.A. = 1$^h$33$^m$24\fs1, Dec. = +30\arcdeg25\arcmin34\farcs8; J2000.0) 
with an arrow.}
\label{fig:m33}
\end{figure*}

\begin{figure*}[ht]
\begin{center}
\plotone{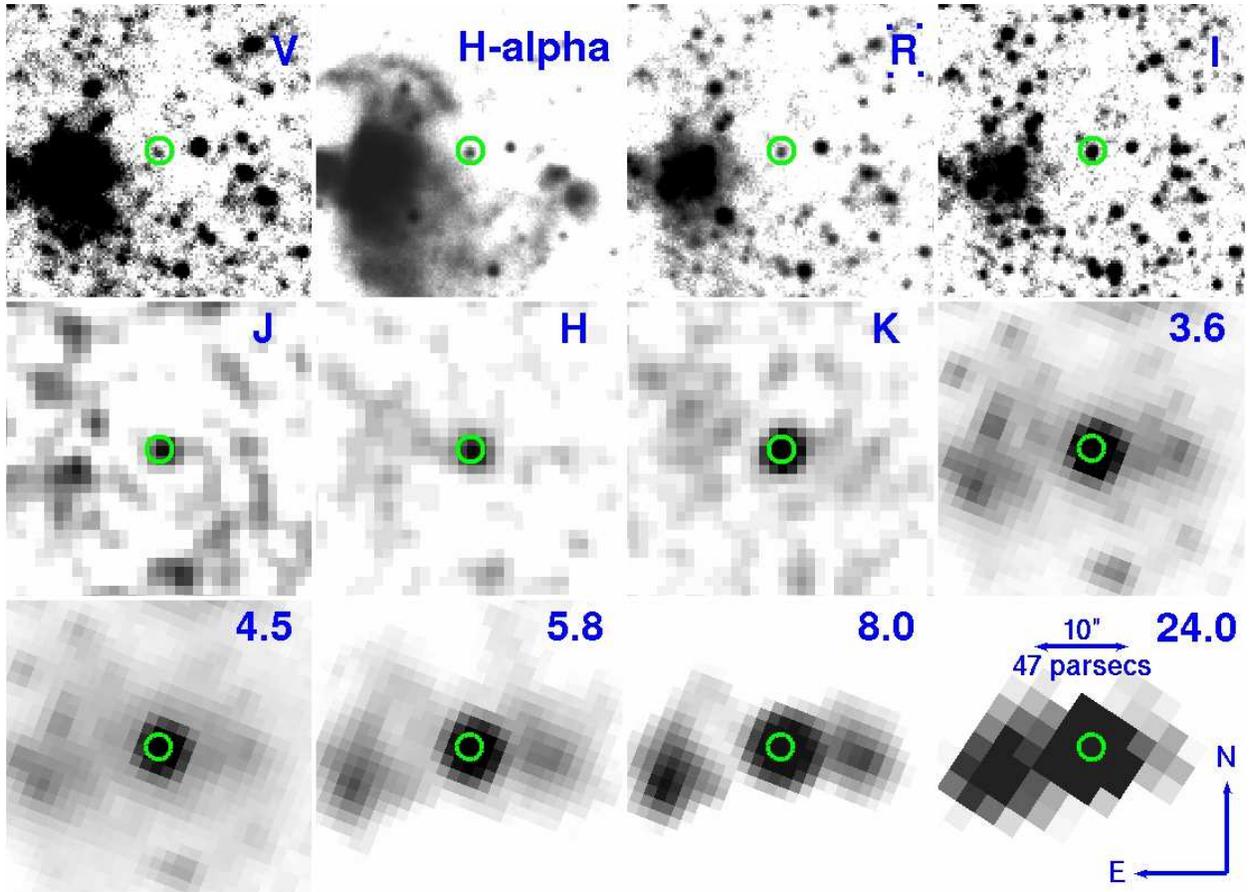}
\end{center}
\caption{Multi-band images of the region around Object~X (marked by the circles). The optical images were taken from the Local Group Survey~\citep{ref:Massey_2006}, the near-IR images were taken from 2MASS~\citep{ref:Skrutskie_2006}, the mid-IR images are from six co-added epochs of the data from~\cite{ref:McQuinn_2007} as processed by \cite{ref:Thompson_2009}, and the MIPS 
24~$\micron$ band image is from the \textit{Spitzer} archive.}
\label{fig:all_bands}
\end{figure*}

Figure~\ref{fig:m33} shows its location in M33, and Figure~\ref{fig:all_bands} shows 
the $V$-band through 24~$\micron$ band images of its surrounding region based on the 
data we describe in Section~\ref{sec:data}. Section~\ref{sec:analysis} presents the 
SED and simple models of it. Section~\ref{sec:discussion} 
discusses the possible explanations of the physical nature of Object~X and its 
implications for understanding the late-stage evolution of the most massive stars.

\section{Data}
\label{sec:data}

The optical photometry ($UBVRI$ and H$\alpha$) was measured from the Local Group Galaxies Survey images \citep{ref:Massey_2006}. The catalog of M33 point sources published by \cite{ref:Massey_2006} does not include Object~X as a point source because it did not satisfy their criterion that it is $>$4 sigma above the background. We used DAOPHOT/ALLSTAR~\citep{ref:Stetson_1992} to identify 
point sources in the $UBVRI$ bands. Object~X is identified as a relatively 
faint and red point source in the $VRI$ bands with a $>3\sigma$ detection. 
However, it is not detected in the $U$ and $B$ bands. We estimated $3\sigma$ 
upper limits on its luminosity in these bands using the APPHOT/PHOT package. 
The measured magnitudes and the limits were transformed to the Vega-calibrated 
system using zero-point offsets determined from the bright stars in 
the \cite{ref:Massey_2006} catalog of M33 point sources. 

The near-IR ($JHK_s$) images were taken from the 2MASS survey~\citep{ref:Skrutskie_2006} and the 
calibrated $JHK_s$ magnitudes were obtained from the 2MASS All-Sky Catalog of Point 
Sources~\citep{ref:Cutri_2003}. 

For the mid-IR photometry, we used the six co-added epochs of data 
from~\cite{ref:McQuinn_2007} as processed and used by~\cite{ref:Thompson_2009}
 and \cite{ref:Khan_2010a}. For the MIPS 24~$\micron$ band, we downloaded the Post-Basic Calibrated 
Data (PBCD) from the \textit{Spitzer} archive (Program~5, PI: Gehrz). Although we examined the MIPS 70~$\micron$ and 
160~$\micron$ images for the target region as well, we were unable to obtain 
reliable flux measurements due to the nearby \ion{H}{2} region
being extremely luminous in these two bands. We measured 
the magnitudes using DAOPHOT/ALLSTAR~\citep{ref:Stetson_1992}. The 
PSF-magnitudes obtained with ALLSTAR were transformed to Vega-calibrated 
magnitudes using aperture corrections derived from bright stars using the 
APPHOT/PHOT package. Figures~\ref{fig:m33_CMD1} 
and \ref{fig:m33_CMD2} show the mid-IR CMDs of M33.

The measured fluxes are reported in Table~1, and $V$-band through 24~$\micron$ band images of 
Object~X are shown in Figure~\ref{fig:all_bands}.

\begin{figure*}[ht]
\begin{center}
\plotone{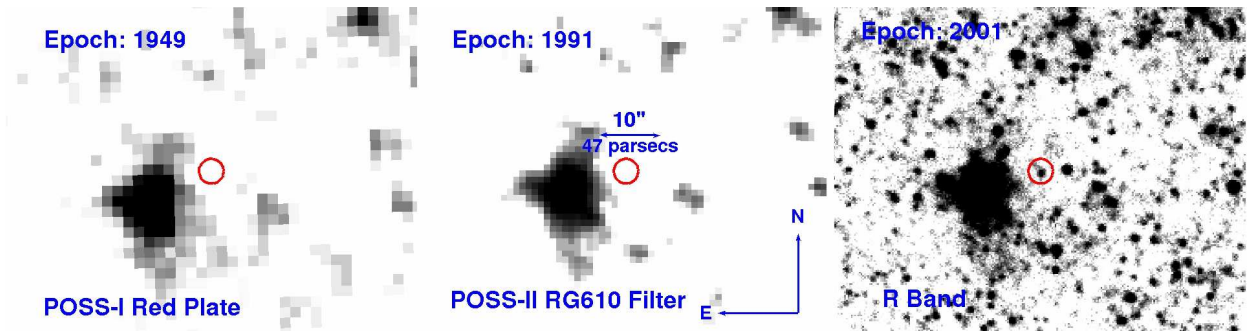}
\end{center}
\caption{Historical images of the region near the location of Object~X (marked by the circles) over the last $\sim60$ years implying that Object~X has remained obscured at least over this period of time.}
\label{fig:history}
\end{figure*}

We also examined the photographic plates from the 48~inch 
Oschin Schmidt Telescope used in the STScI
Digitized Sky Survey. The first image (epoch 1949) was a 
POSS-I Red Plate, while the second (epoch 1991) was a POSS-II 
RG610 filter. Both of these are roughly comparable to the modern standard 
$R$-band. Neither image shows an optical source at the location of Object~X. 
These images are much shallower than the Local Group Survey~\citep{ref:Massey_2006}. 
Based on the faintest $R$-band USNO stars in this region and calibrating them by 
their counterparts in~\cite{ref:Massey_2006}, we estimate an upper limit of 
R$\gtrsim20.0$ magnitude for Object~X in 1949. This is consistent with our 
detection of Object~X at R$\simeq21.6$ (epoch 2001). The epoch 1991 image also does 
not show a source, but we were unable to converge on a self-consistent calibration 
for the region, and so will not discuss this image further. Figure~\ref{fig:history} 
shows these historical images of this region.

\begin{figure*}[ht]
\begin{center}
\plotone{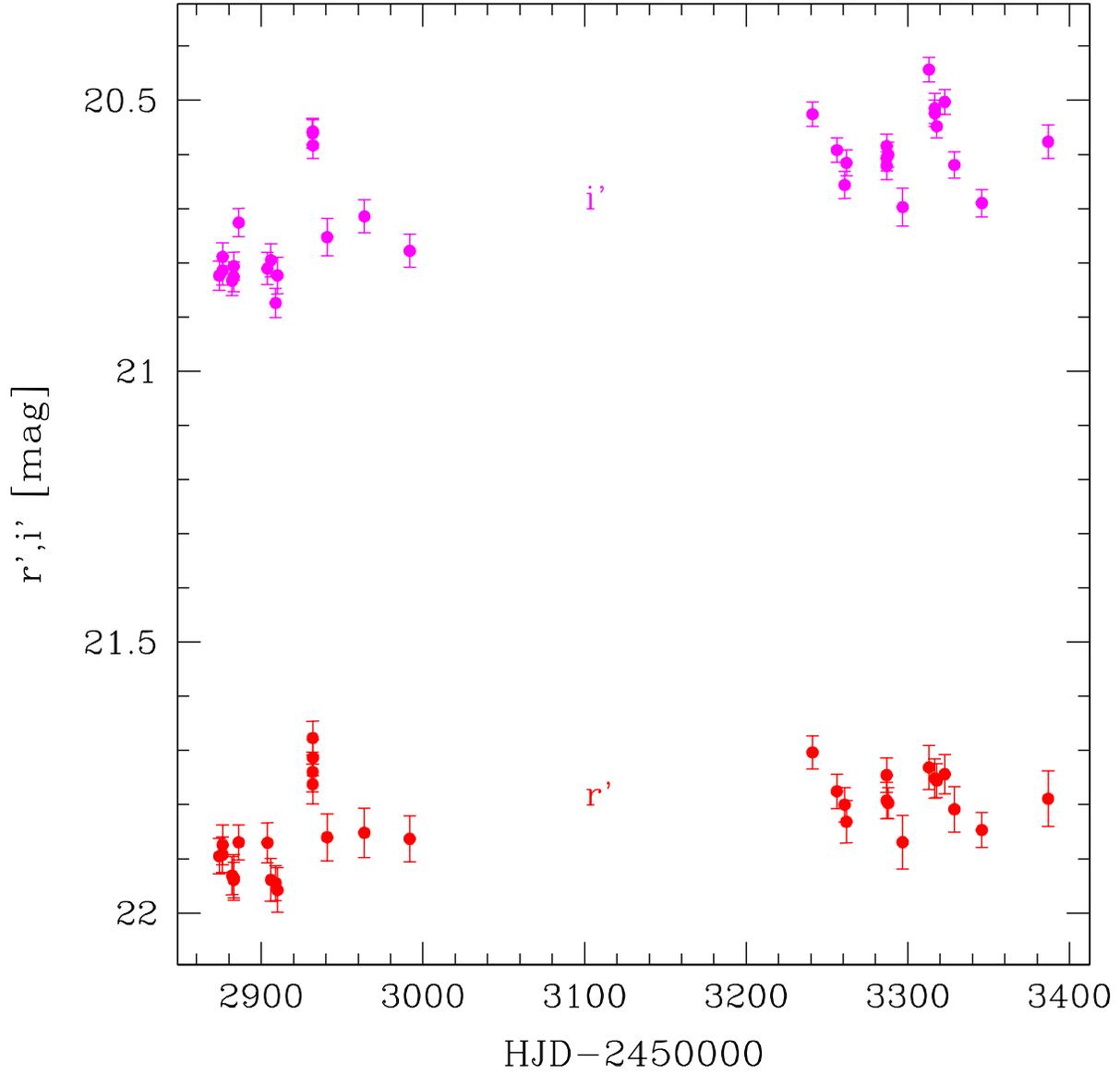}
\end{center}
\caption{Optical variability of Object~X in the \textit{r'} and \textit{i'} bands from~\cite{ref:Hartman_2006}.
Object~X shows correlated variability of about $\sim$0.4 magnitude (fractional variability of $\sim$45\%) in both bands.}
\label{fig:opt_lc}
\end{figure*}

Object~X is identified as an optically variable point source in the 
Canada-France-Hawaii Telescope (CFHT) photometric survey of M33~\citep{ref:Hartman_2006}. 
Figure~\ref{fig:opt_lc} shows the $r'$ and $i'$ band lightcurves of Object~X 
from this survey. The correlated short-term variability of $\sim0.4$~magnitude
(fractional variability of $\sim$45\%),
definitively indicates that it is a single stellar object rather than 
multiple objects blended together. If we were to dilute the source with 
additional non-variable sources of comparable total luminosity, the observed 
$\sim0.4$~magnitude variability would be unphysical. This is significant 
given the lack of archival HST images for this location. 

\begin{figure*}[ht]
\begin{center}
\plotone{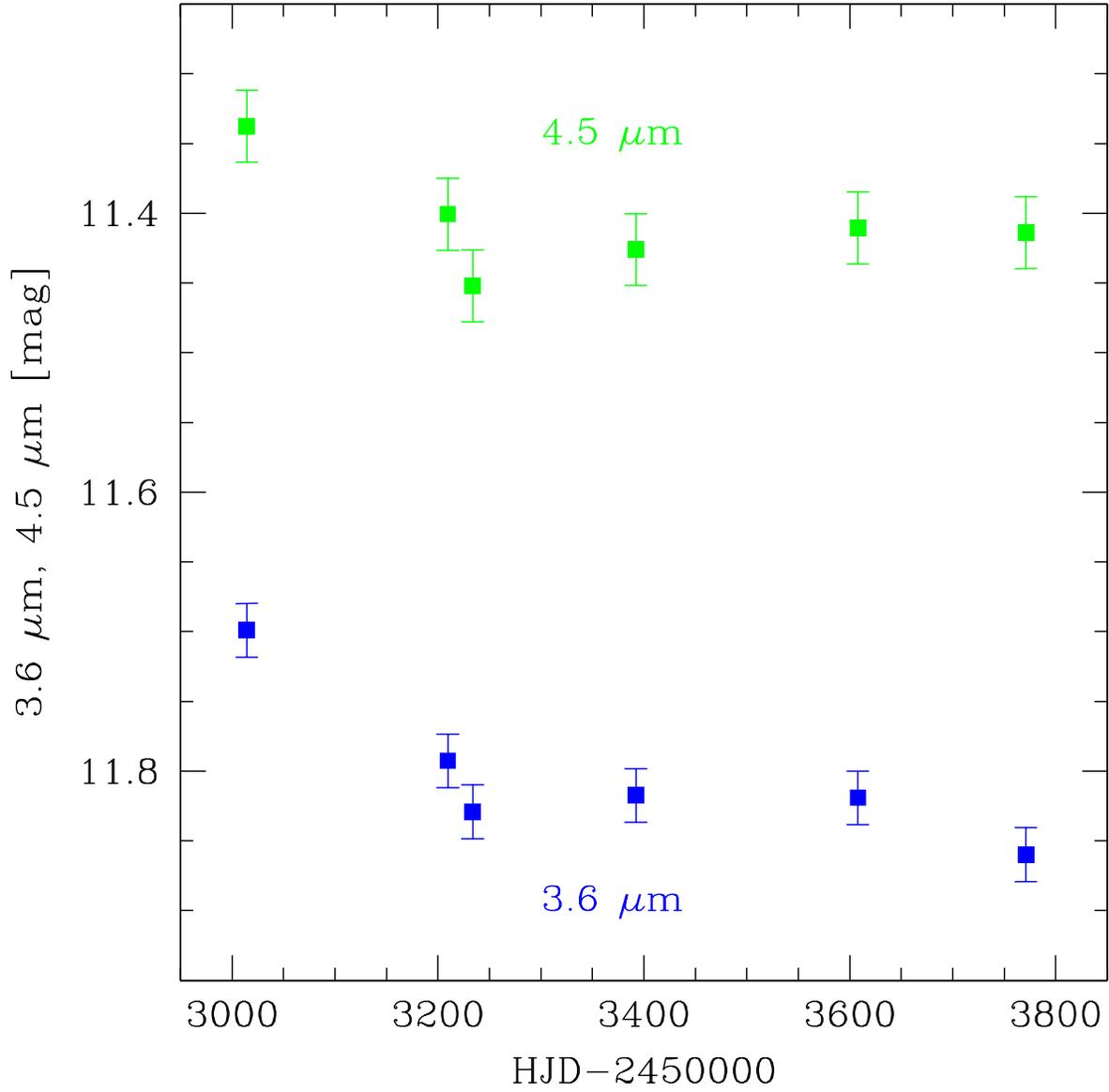}
\end{center}
\caption{Mid-IR variability of Object~X in the 3.6~$\micron$ and 4.5~$\micron$ IRAC bands. It shows correlated variability of about $\sim$0.15 magnitude (fractional variability of $\sim$15\%) in both bands.}
\label{fig:mir_lc}
\end{figure*}

\cite{ref:McQuinn_2007} collected six epochs over two years and identified Object~X as a variable source in the mid-IR,
although for this paper, we use the mid-IR lightcurves generated by \cite{ref:Thompson_2009}. 
Figure~\ref{fig:mir_lc} shows the 3.6~$\micron$ and 4.5~$\micron$ band 
lightcurves. Object~X shows correlated variability of about 
$\sim$0.15 magnitude in both bands (fractional variability of 
$\sim$15\%). This suggests that the dust opacity and geometry surrounding Object~X 
are not undergoing any rapid evolution.

\section{Analysis}
\label{sec:analysis}

Figure~\ref{fig:sed} shows the spectral energy distribution (SED) of Object~X. 
The low optical luminosity means that Object~X is unlikely to be 
a YSO embedded in a dusty disk, and if we model the SED using the YSO models of 
\cite{ref:Robitaille_2006}, we find no good, or even plausible fits\footnote{We fit the SED using the online tool~\citep{ref:Robitaille_2007} at\newline \texttt{http://caravan.astro.wisc.edu/protostars/sedfitter.php}.}. In essence, $\lesssim$1\% of the luminosity is emitted in the optical, while even edge-on YSO disk models tend to scatter more optical light into the observer's line of sight.

The alternate is that we are examining a self-obscured star. For this case,
we fit the SED of the source using DUSTY~\citep{ref:Ivezic_1997,ref:Ivezic_1999,ref:Elitzur_2001} 
to model the radiation transfer through a spherical dusty medium surrounding a star. We considered models
using either the graphitic or silicate dust models of \cite{ref:Draine_1984}. We distributed the dust
either as a shell, with ratios of outer to inner radii of either 2:1 or 4:1, or with the standard
DUSTY wind model. The density distributions inside the shells were assumed to be $\rho \propto 1/r^2$.
The models are defined by the stellar luminosity, $L_*$, stellar temperature, $T_*$, the V-band
optical depth $\tau_V$, and the dust temperature at the inner edge of the dust distribution, $T_d$.
The stellar luminosity and the dust temperature together determine the radius of the inner edge of the dust distribution.
To fit the data we tabulated the DUSTY models on a grid of stellar temperatures, optical depths
and dust temperatures.

\begin{figure}[t]
\begin{center}
\includegraphics[width=120mm]{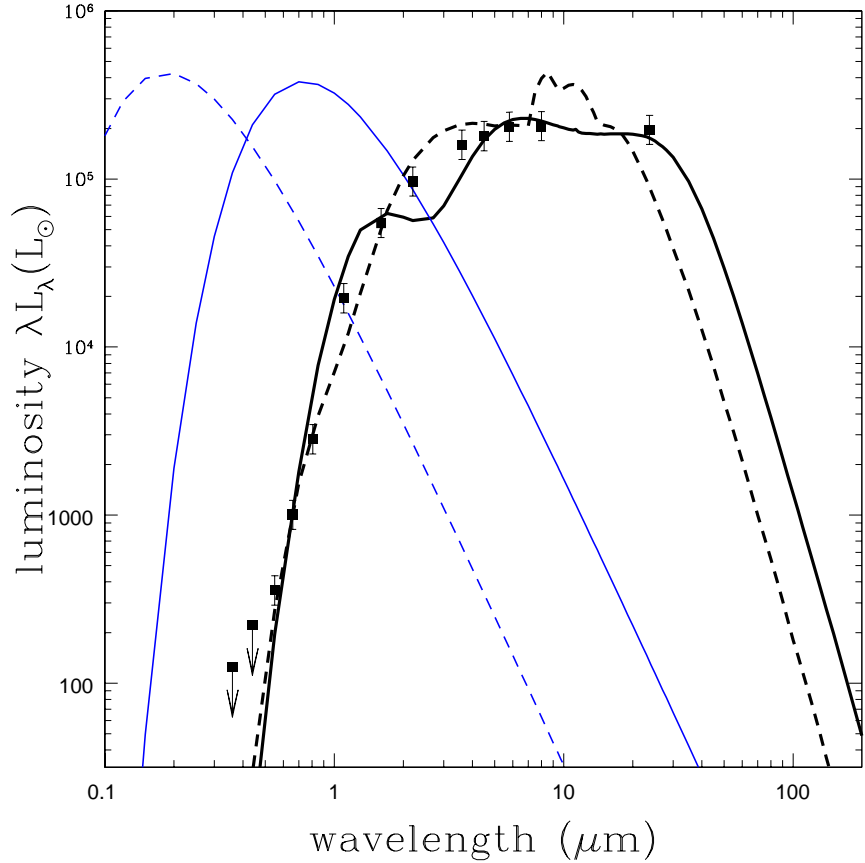}
\end{center}
\caption{The results of DUSTY fits to Object~X SED (black points). The heavier 
solid (dashed) lines show the model DUSTY SEDs for the source for graphitic 
(silicate) dusts. The lighter solid (dashed) lines show the black-body stellar 
SEDs. The graphitic model has $T_*=5000$~K and $L_* = 10^{5.7}L_\odot$ with 
$\tau_V=8.5$ and an inner edge dust temperature of $500$~K corresponding to an 
inner shell radius of $2.6 \times 10^{16}$~cm. The silicate model has 
$T_*=20000$~K and $L_* = 10^{5.8} L_\odot$ with $\tau_V=11.5$ and an inner edge 
dust temperature of $1200$~K corresponding to an inner shell radius of 
$6.4\times 10^{15}$~cm. These models have a 2:1 ratio between their inner and 
outer radii, but the 4:1 and wind models (not shown) look very similar.}
\label{fig:dusty}
\end{figure}

We find that two dust types that can fit the SED equally well
but with very different preferred stellar
temperatures. The graphitic models prefer cool stars, $T_* \simeq 5000$~K, while the
silicate models prefer hot stars, $T_* \simeq 20000$~K. The stellar luminosity is
$L_* \simeq 10^{5.6} L_\odot$ to $10^{5.8} L_\odot$ with the silicate models favoring
the somewhat higher luminosities (see Figure~\ref{fig:dusty}). The three different dust geometries produce similarly
good statistical fits. For the fits we assumed minimum flux uncertainties of 20\%
and included the optical upper limits as constraints, leading to $\chi^2 \simeq 43$ for
$N_{dof}=9$ for the best fits. Each model was a modestly poor match to some part of
the SED, but this is not very surprising given our use of a simple spherical model
and a discrete sampling of its parameter space. A clear detection (or not) of the 
silicate dust feature near 10~$\micron$ would determine the dust composition. 
The typical optical depths and
inner/outer dust temperatures were $\tau_V \simeq 8.5$ and $T_d\simeq 500$/$150$~K
for the graphitic models, and $\tau_V \simeq 11.5$ and $T_D \simeq 1200$/$300$~K
for the silicate models. Much of this difference is created by the higher scattering opacities 
of the silicate dust. The inner edge of the dust lies at R$_{in}\simeq$10$^{16.3}$~cm 
for the graphitic models and R$_{in}\simeq$10$^{15.8}$~cm from the silicate models.

Based on the CO and \ion{H}{1} data from~\citep{ref:Gratier_2010}, the estimated column
density at the location of Object X is $\sim 7 \times 10^{21}$ atoms/cm$^2$
(P.~Gratier, private communication), where the higher number comes
from using the CO to estimate the amount of H$_2$. For a standard
Galactic dust to gas ratio of $E(B-V)=1.7\times10^{-22}$~mag cm$^2$/atom~\citep{ref:Bohlin_1978}, 
this implies a maximum foreground extinction in M33
of roughly $E(B-V)\simeq$ 1.2 magnitude. Adding this to the Galactic
extinction changes the DUSTY models very little, modestly reducing
the optical depths ($\Delta \tau \sim 3-4$) and slightly raising the
stellar luminosity (by $\sim$20\%).

If we search the Padova stellar models~\citep{ref:Marigo_2008} for stars with such
luminosities, there are examples with both the high and low stellar temperatures,
although the higher temperature stars of the silicate models are preferred. The
stellar masses are always $\gtrsim 30 M_\odot$, but without an independent
constraint on the stellar temperature it is difficult to say more. In many,
but not all cases, the stars have lost significant fractions of their initial
masses.

\begin{figure*}[ht]
\begin{center}
\plotone{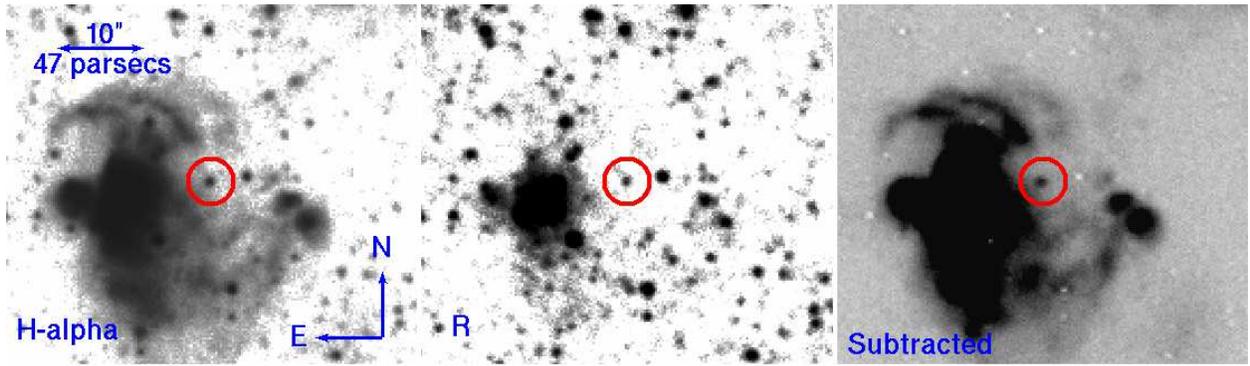}
\end{center}
\caption{The H$\alpha$, R-band, and H$\alpha-$R subtracted images of the region near Object~X. In the subtracted image, almost all the objects vanish, leaving behind the nearby \ion{H}{2} region, Object~X, and a small number of additional, mostly extended, H$\alpha$ sources.}
\label{fig:Ha-R}
\end{figure*}

We also note that while stars ordinarily do not have significant H$\alpha$ emission, 
Object~X is a strong H$\alpha$ source. We determine this by using difference imaging 
methods to match the H$\alpha$ image to the 
flux scale and PSF structure of the $R$-band image and 
then subtract. All the normal stars disappear to leave us only with the stars 
having significant H$\alpha$ emission. In practice, we use the ISIS image 
subtraction software package~\citep{ref:Alard_1998,ref:Alard_2000} following the procedures 
of~\cite{ref:Khan_2010a}. Figure~\ref{fig:Ha-R} shows the result of this procedure for 
the region around Object~X. It is clear from the subtracted image that 
Object~X is a strong H$\alpha$ emission source.

\section{Discussion}
\label{sec:discussion}

An evolved star can cloak itself in dust through two broad mechanisms,
winds and mass ejection. The classic examples of wind obscured stars
are the AGB stars. In the most extreme cases, such as the
progenitors of SN~2008S and the 2008 NGC~300 transient, the wind
can be optically thick even in the mid-IR~\citep{ref:Prieto_2008a}.
Shells formed by impulsive mass ejections are seen around many
evolved massive stars~\citep[e.g.,][]{ref:Humphreys_1994,ref:Humphreys_1997,ref:Smith_2010,ref:Gvaramadze_2010,ref:Wachter_2010}, and they are distinguished by the
frequency of the ejections, their mass, and their velocities.
The most famous example is the ``Great Eruption'' of $\eta$ Carina
in the 19th century, which ejected $\sim 10M_\odot$ of material
at velocity $\sim 600$~kms$^{-1}$~\citep{ref:Humphreys_1994}. At the other end of the velocity
spectrum are the OH/IR stars such as IRC+10420, where the ejection
velocities are closer to $\sim 50$~kms$^{-1}$~\citep{ref:Tiffany_2010}.

Our best constraint on these possibilities comes from the time variability
of the source. The absence of the source in 1949 means that the
star has been obscured for at least 60 years, but the SED models
require fairly hot dust close to the star. If we assume that
the material obscuring the star today is the same as that 
obscuring the star in 1949, we get characteristic velocities
of the order of $\sim 114$~kms$^{-1}$ and $\sim 36$~kms$^{-1}$ for the graphitic
and silicate models, respectively. The outer edge then requires
a velocity twice as large for a shell with a 2:1 ratio between
the inner and outer edges. These velocity scales favor the slower
ejections of the OH/IR stars over the higher velocities of
the giant LBV eruptions like $\eta$ Carina. The significant optical
and mid-IR variability then favors a patchy evolving ejecta
over a steady wind.  The extended period of the obscuration
comes close to requiring multiple ejections, approaching the
limit of an unsteady wind. It would be interesting to determine
if the source was visible in still older archival plates, if any exist, since
it would be a relatively bright $\hbox{R} \simeq 16$~mag source
without the dust.

\begin{figure*}[t]
\begin{center}
\includegraphics[width=120mm]{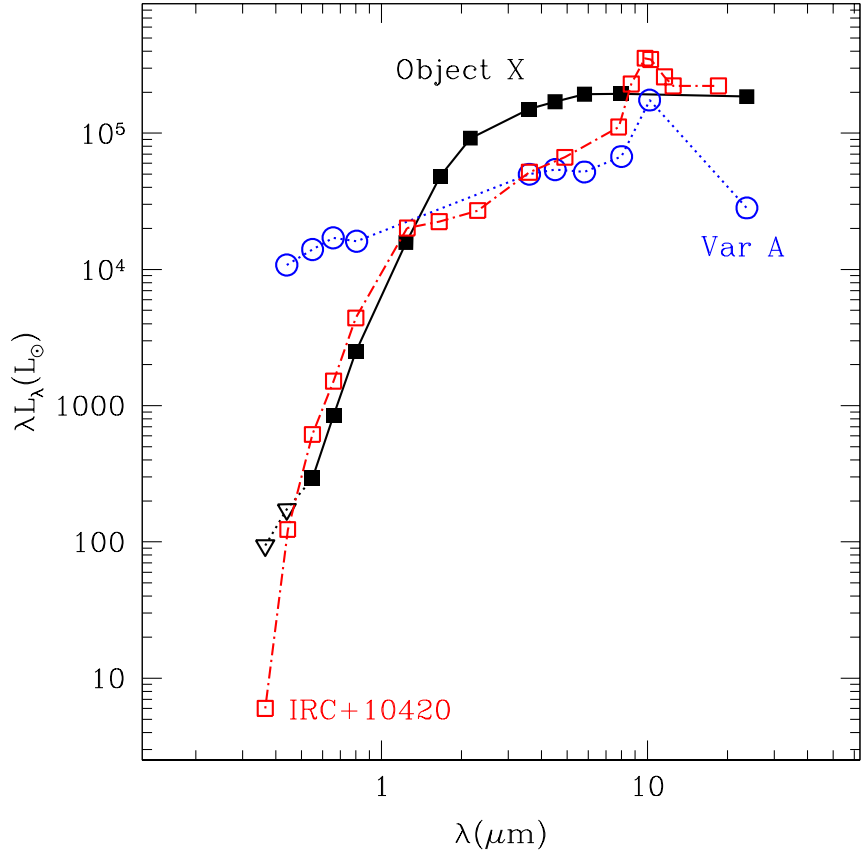}
\end{center}
\caption{The SED of Object~X as compared to that of IRC+10420~\citep{ref:Jones_1993,ref:Humphreys_1997} and Var~A~\citep{ref:Humphreys_2006}. Both comparison SEDs have been constructed using data from multiple epochs (IRC+10420: $UB$ from 1972, $VRI$ from 1992, the rest from 1996; Var~A: optical from 2000-01, near-IR from 1997, the rest from 2004-05) and these sources are known to be variable. We also note the large amount of extinction toward IRC+10420~\citep[$A_V \simeq 6$,][]{ref:Jones_1993}. correcting for this extinction would make it more similar to Var~A than Object~X. However, it is uncertain how much of the extincting material is associated with that star rather than simply being along the line of sight, and therefore we do not correct for the extinction here.}
\label{fig:sed}
\end{figure*}

Overall, Object~X seems to most closely resemble the cool,
hypergiant, stars such as IRC+10420 and Var~A. The three sources have comparable bolometric
luminosities of $\sim10^{5.7}L_\odot$, although their
SEDs, shown in Figure~\ref{fig:sed}, differ because Object~X is
more obscured. In the case of
IRC+10420, the star is surrounded by a series of shells
and seems to have undergone a period of high mass loss
for the last $\sim 600$~years~\citep{ref:Humphreys_1997} that ended 50-100 years ago
leading to its current optical brightening~\citep{ref:Blocker_1999}. Var~A,
on the other hand, seems to have had a brief period
of high mass loss rates over the last $\sim 50$ years
and is now re-emerging in the optical~\citep{ref:Humphreys_2006}. Object~X would
seem to be intermediate, requiring a longer period of
heavy mass loss than Var~A, but perhaps less than
IRC+10420. If Object~X is 
a true analog to IRC+10420 or Var~A, then it may conceivably emerge from its current self-obscured 
state over the next few decades since the optical depth of an expanding shell drops as 
$\tau \propto t^{-2}$. We cannot make this statement with any degree of 
certainty, as each of these objects are going through tumultuous evolutionary phases 
that are both unique and poorly understood. Follow-up observations 
of this source at all wavelengths should begin to reveal its long term evolution and may 
provide us with unexpected surprises. 
Deep optical and IR spectra of Object~X will enable to us to determine the spectral type of the central star
and place it on an H-R diagram.

Stars like Object~X are extremely rare. As a unique star in a brief yet eventful evolutionary state, 
we encourage further study of this interesting object. Furthermore, there is mounting evidence that 
mass loss from massive stars may be dominated by impulsive transients rather than steady winds~\citep[e.g.,][]{ref:Smith_2006}, 
particularly with the downward revision of mass-loss rates 
in recent years~\citep[e.g.,][]{ref:Fullerton_2006}. If so, the period in which mass loss is most important will also tend to 
be the period when the star is most obscured. Surveys of massive stars such as those by \cite{ref:Massey_2006} 
and \cite{ref:Bonanos_2009,ref:Bonanos_2010}, which focus on bright optical sources, will 
miss Object~X and similar sources that may best probe the frequency and duration of these 
mass loss episodes as well as the amount of ejected mass. Characterizing these stars and the 
amount of mass loss by massive stars clearly requires systematic surveys in the mid-IR as 
well as the optical.

\acknowledgments
We thank the referee, J.~L.~Prieto, P.~Gratier and R.~Humphreys for their useful advice, 
T.~A.Thompson, B.~Metzger and K.~Sellgren for helpful discussions, and D.~Szczygiel 
for helping us analyze the MIPS data.
We extend our gratitude to the SINGS Legacy Survey and LVL Survey for making 
their data publicly available. This research has made use of NED, which is 
operated by the JPL and Caltech, under contract with NASA and the HEASARC 
Online Service, provided by NASA's GSFC. This research has made use of 
photographic data of the National Geographic Society -- Palomar Observatory Sky 
Survey (NGS-POSS) obtained using the Oschin Telescope on Palomar Mountain. The 
NGS-POSS was funded by a grant from the National Geographic Society to the 
California Institute of Technology. The Digitized Sky Survey was produced at 
the Space Telescope Science Institute under US Government grant NAG W-2166. RK 
and KZS are supported in part by NSF grant AST-0707982. KZS and CSK are 
supported in part by NSF grant AST-0908816. AZB acknowledges research and travel 
support from the European Commission Framework Program Seven under a Marie Curie 
International Reintegration Grant.

\begin{table}[p]
\begin{center}
\label{table:data}
\begin{tabular}{ccccc}
\hline 
\hline
\\
\multicolumn{1}{c}{Band} &
\multicolumn{1}{c}{Magnitude} &
\multicolumn{1}{c}{Flux} &
\multicolumn{1}{c}{Luminosity} &
\multicolumn{1}{c}{Year}
\\
\multicolumn{1}{c}{} &
\multicolumn{1}{c}{} &
\multicolumn{1}{c}{(mJy)} &
\multicolumn{1}{c}{(log($\lambda L_{\lambda} / L_{\odot}$))} &
\multicolumn{1}{c}{}
\\
\hline
\hline
\\
$U$ & $\gtrsim24.1$ & $\lesssim4.20 \times 10^{-4}$ & $\lesssim$1.98 & 2001\\
$B$ & $\gtrsim24.2$ & $\lesssim9.39 \times 10^{-4}$ & $\lesssim$2.24 & 2001\\
$V$ & $23.15\pm0.13$ & $1.97 \times 10^{-3}$ & 2.467 & 2001\\
$R$ & $21.61\pm0.04$ & $6.83 \times 10^{-3}$ & 2.927 & 2001\\
$I$ & $19.99\pm0.02$ & $2.43 \times 10^{-2}$ & 3.395 & 2001\\
$J$ & $17.07\pm0.17$ & $2.37 \times 10^{-1}$ & 4.197 & 2001\\
$H$ & $15.04\pm0.09$ & $9.83 \times 10^{-1}$ & 4.684 & 2001\\
$K_s$ & $13.60\pm0.04$ & $2.42$ & 4.963 & 2001\\
3.6~$\micron$ & $11.57\pm0.05$ & $6.62$ & 5.177 & 2004\\
4.5~$\micron$ & $10.71\pm0.05$ & $9.34$ & 5.230 & 2004\\
5.8~$\micron$ & $9.81\pm0.02$ & $13.7$ & 5.286 & 2004\\
8.0~$\micron$ & $8.83\pm0.01$ & $19.1$ & 5.290 & 2004\\
24.0~$\micron$ & $5.31\pm\sim0.1$ & $53.7$ & 5.268 & 2005\\
\hline
\hline
\end{tabular}
\end{center}
\caption{Photometry of Object~X. For computing the luminosity we used a distance of 0.96~Mpc~\citep{ref:Bonanos_2006} and corrected for Galactic extinction of $E(B-V)=0.049$ \citep{ref:Schlegel_1998}. The $UBVRI$ images are from \cite{ref:Massey_2006}, the $JHK_s$ photometry is from \cite{ref:Cutri_2003}, the IRAC bands photometry is from \cite{ref:Thompson_2009}, and the MIPS 24~$\micron$ band image from the \textit{Spitzer} archive (Program 5, PI:Gehrz).}
\end{table}

\bibliographystyle{apj}

\bibliography{bibliography}

\begin{thebibliography}{37}
\expandafter\ifx\csname natexlab\endcsname\relax\def\natexlab#1{#1}\fi

\bibitem[{{Alard}(2000)}]{ref:Alard_2000}
{Alard}, C. 2000, \aaps, 144, 363

\bibitem[{{Alard} \& {Lupton}(1998)}]{ref:Alard_1998}
{Alard}, C. \& {Lupton}, R.~H. 1998, \apj, 503, 325

\bibitem[{{Bl{\"o}cker} {et~al.}(1999){Bl{\"o}cker}, {Balega}, {Hofmann},
  {Lichtenth{\"a}ler}, {Osterbart}, \& {Weigelt}}]{ref:Blocker_1999}
{Bl{\"o}cker}, T., {Balega}, Y., {Hofmann}, K., {Lichtenth{\"a}ler}, J.,
  {Osterbart}, R., \& {Weigelt}, G. 1999, \aap, 348, 805

\bibitem[{{Bohlin} {et~al.}(1978){Bohlin}, {Savage}, \&
  {Drake}}]{ref:Bohlin_1978}
{Bohlin}, R.~C., {Savage}, B.~D., \& {Drake}, J.~F. 1978, \apj, 224, 132

\bibitem[{{Bonanos} {et~al.}(2006){Bonanos}, {Stanek}, {Kudritzki}, {Macri},
  {Sasselov}, {Kaluzny}, {Stetson}, {Bersier}, {Bresolin}, {Matheson},
  {Mochejska}, {Przybilla}, {Szentgyorgyi}, {Tonry}, \&
  {Torres}}]{ref:Bonanos_2006}
{Bonanos}, A.~Z., {Stanek}, K.~Z., {Kudritzki}, R.~P., {Macri}, L.~M.,
  {Sasselov}, D.~D., {Kaluzny}, J., {Stetson}, P.~B., {Bersier}, D.,
  {Bresolin}, F., {Matheson}, T., {Mochejska}, B.~J., {Przybilla}, N.,
  {Szentgyorgyi}, A.~H., {Tonry}, J., \& {Torres}, G. 2006, ApJ, 652, 313

\bibitem[{{Bonanos} {et~al.}(2009)}]{ref:Bonanos_2009}
{Bonanos}, A.~Z. {et~al.} 2009, AJ, 138, 1003

\bibitem[{{Bonanos} {et~al.}(2010)}]{ref:Bonanos_2010}
---. 2010, AJ, 140, 416

\bibitem[{{Cutri} {et~al.}(2003)}]{ref:Cutri_2003}
{Cutri}, R.~M. {et~al.} 2003, {2MASS All Sky Catalog of point sources.}, ed.
  R.~M. Cutri {et~al.}

\bibitem[{{Draine} \& {Lee}(1984)}]{ref:Draine_1984}
{Draine}, B.~T. \& {Lee}, H.~M. 1984, \apj, 285, 89

\bibitem[{{Elitzur} \& {Ivezi{\'c}}(2001)}]{ref:Elitzur_2001}
{Elitzur}, M. \& {Ivezi{\'c}}, {\v Z}. 2001, \mnras, 327, 403

\bibitem[{{Fullerton} {et~al.}(2006){Fullerton}, {Massa}, \&
  {Prinja}}]{ref:Fullerton_2006}
{Fullerton}, A.~W., {Massa}, D.~L., \& {Prinja}, R.~K. 2006, \apj, 637, 1025

\bibitem[{{Gal-Yam} {et~al.}(2007)}]{ref:GalYam_2007}
{Gal-Yam}, A. {et~al.} 2007, \apj, 656, 372

\bibitem[{{Gratier} {et~al.}(2010)}]{ref:Gratier_2010}
{Gratier}, P. {et~al.} 2010, \aap, 522, A3+

\bibitem[{{Gvaramadze} {et~al.}(2010){Gvaramadze}, {Kniazev}, \&
  {Fabrika}}]{ref:Gvaramadze_2010}
{Gvaramadze}, V.~V., {Kniazev}, A.~Y., \& {Fabrika}, S. 2010, \mnras, 405, 1047

\bibitem[{{Hartman} {et~al.}(2006){Hartman}, {Bersier}, {Stanek}, {Beaulieu},
  {Kaluzny}, {Marquette}, {Stetson}, \&
  {Schwarzenberg-Czerny}}]{ref:Hartman_2006}
{Hartman}, J.~D., {Bersier}, D., {Stanek}, K.~Z., {Beaulieu}, J., {Kaluzny},
  J., {Marquette}, J., {Stetson}, P.~B., \& {Schwarzenberg-Czerny}, A. 2006,
  MNRAS, 371, 1405

\bibitem[{{Humphreys} \& {Davidson}(1994)}]{ref:Humphreys_1994}
{Humphreys}, R.~M. \& {Davidson}, K. 1994, PASP, 106, 1025

\bibitem[{{Humphreys} {et~al.}(1997)}]{ref:Humphreys_1997}
{Humphreys}, R.~M. {et~al.} 1997, \aj, 114, 2778

\bibitem[{{Humphreys} {et~al.}(2006)}]{ref:Humphreys_2006}
---. 2006, AJ, 131, 2105

\bibitem[{{Ivezic} \& {Elitzur}(1997)}]{ref:Ivezic_1997}
{Ivezic}, Z. \& {Elitzur}, M. 1997, \mnras, 287, 799

\bibitem[{{Ivezic} {et~al.}(1999){Ivezic}, {Nenkova}, \&
  {Elitzur}}]{ref:Ivezic_1999}
{Ivezic}, Z., {Nenkova}, M., \& {Elitzur}, M. 1999, arXiv:astro-ph/9910475

\bibitem[{{Jones} {et~al.}(1993)}]{ref:Jones_1993}
{Jones}, T.~J. {et~al.} 1993, \apj, 411, 323

\bibitem[{{Khan} {et~al.}(2010){Khan}, {Stanek}, {Prieto}, {Kochanek},
  {Thompson}, \& {Beacom}}]{ref:Khan_2010a}
{Khan}, R., {Stanek}, K.~Z., {Prieto}, J.~L., {Kochanek}, C.~S., {Thompson},
  T.~A., \& {Beacom}, J.~F. 2010, ApJ, 715, 1094

\bibitem[{{Marigo} {et~al.}(2008){Marigo}, {Girardi}, {Bressan}, {Groenewegen},
  {Silva}, \& {Granato}}]{ref:Marigo_2008}
{Marigo}, P., {Girardi}, L., {Bressan}, A., {Groenewegen}, M.~A.~T., {Silva},
  L., \& {Granato}, G.~L. 2008, \aap, 482, 883

\bibitem[{{Massey} {et~al.}(2006){Massey}, {Olsen}, {Hodge}, {Strong},
  {Jacoby}, {Schlingman}, \& {Smith}}]{ref:Massey_2006}
{Massey}, P., {Olsen}, K.~A.~G., {Hodge}, P.~W., {Strong}, S.~B., {Jacoby},
  G.~H., {Schlingman}, W., \& {Smith}, R.~C. 2006, AJ, 131, 2478

\bibitem[{{McQuinn} {et~al.}(2007)}]{ref:McQuinn_2007}
{McQuinn}, K.~B.~W. {et~al.} 2007, ApJ, 664, 850

\bibitem[{{Prieto} {et~al.}(2008)}]{ref:Prieto_2008a}
{Prieto}, J.~L. {et~al.} 2008, ApJ Letters, 681, L9

\bibitem[{{Robitaille} {et~al.}(2007){Robitaille}, {Whitney}, {Indebetouw}, \&
  {Wood}}]{ref:Robitaille_2007}
{Robitaille}, T.~P., {Whitney}, B.~A., {Indebetouw}, R., \& {Wood}, K. 2007,
  ApJ Supplement, 169, 328

\bibitem[{{Robitaille} {et~al.}(2006){Robitaille}, {Whitney}, {Indebetouw},
  {Wood}, \& {Denzmore}}]{ref:Robitaille_2006}
{Robitaille}, T.~P., {Whitney}, B.~A., {Indebetouw}, R., {Wood}, K., \&
  {Denzmore}, P. 2006, ApJ Supplement, 167, 256

\bibitem[{{Schlegel} {et~al.}(1998){Schlegel}, {Finkbeiner}, \&
  {Davis}}]{ref:Schlegel_1998}
{Schlegel}, D.~J., {Finkbeiner}, D.~P., \& {Davis}, M. 1998, \apj, 500, 525

\bibitem[{{Skrutskie} {et~al.}(2006)}]{ref:Skrutskie_2006}
{Skrutskie}, M.~F. {et~al.} 2006, AJ, 131, 1163

\bibitem[{{Smith} {et~al.}(2008){Smith}, {Chornock}, {Li}, {Ganeshalingam},
  {Silverman}, {Foley}, {Filippenko}, \& {Barth}}]{ref:Smith_2008}
{Smith}, N., {Chornock}, R., {Li}, W., {Ganeshalingam}, M., {Silverman}, J.~M.,
  {Foley}, R.~J., {Filippenko}, A.~V., \& {Barth}, A.~J. 2008, \apj, 686, 467

\bibitem[{{Smith} \& {Frew}(2010)}]{ref:Smith_2010}
{Smith}, N. \& {Frew}, D.~J. 2010, arXiv:astro-ph/1010.3719

\bibitem[{{Smith} \& {Owocki}(2006)}]{ref:Smith_2006}
{Smith}, N. \& {Owocki}, S.~P. 2006, \apjl, 645, L45

\bibitem[{{Stetson}(1992)}]{ref:Stetson_1992}
{Stetson}, P.~B. 1992, 25, 297

\bibitem[{{Thompson} {et~al.}(2009){Thompson}, {Prieto}, {Stanek}, {Kistler},
  {Beacom}, \& {Kochanek}}]{ref:Thompson_2009}
{Thompson}, T.~A., {Prieto}, J.~L., {Stanek}, K.~Z., {Kistler}, M.~D.,
  {Beacom}, J.~F., \& {Kochanek}, C.~S. 2009, ApJ, 705, 1364

\bibitem[{{Tiffany} {et~al.}(2010){Tiffany}, {Humphreys}, {Jones}, \&
  {Davidson}}]{ref:Tiffany_2010}
{Tiffany}, C., {Humphreys}, R.~M., {Jones}, T.~J., \& {Davidson}, K. 2010, \aj,
  140, 339

\bibitem[{{Wachter} {et~al.}(2010){Wachter}, {Mauerhan}, {Van Dyk}, {Hoard},
  {Kafka}, \& {Morris}}]{ref:Wachter_2010}
{Wachter}, S., {Mauerhan}, J.~C., {Van Dyk}, S.~D., {Hoard}, D.~W., {Kafka},
  S., \& {Morris}, P.~W. 2010, \aj, 139, 2330

\end{thebibliography}

\end{document}